\documentclass{article}

\pdfoutput=1
\usepackage{graphicx,amsmath,amssymb,epsf} \usepackage{latexsym,bm,
  slashed} \usepackage{xcolor} \definecolor{dark}{rgb}{0.10,0.2,0.3}
\definecolor{magenta}{rgb}{0.7,0.1,0.3}
\definecolor{purpure}{rgb}{0.5,0.15,0.3}
\usepackage[font=small,format=plain,labelfont=bf,up,textfont=it,up]{caption}
\usepackage{hyperref, cite} \hypersetup{colorlinks, citecolor=blue,
  filecolor=blue, linkcolor=magenta,
  urlcolor=purpure,hyperfootnotes=true,pdftex} 

\usepackage{subcaption}

\title{\bf On the conformal spin dependence of the\\ perturbative QCD vacuum singularity} 
\author{Grigorios Chachamis$^1$ and Agust{\' \i}n Sabio Vera$^{2,3}$\\ 
\\
\small $^1$ Laborat{\' o}rio de Instrumenta\c{c}{\~ a}o e F{\' \i}sica Experimental de Part{\' \i}culas (LIP),\\
\small Av. Prof. Gama Pinto, 2, P-1649-003 Lisboa, Portugal.\\
\small $^2$ Instituto de F{\'\i}sica Te{\' o}rica UAM/CSIC, c/ Nicol{\' a}s Cabrera 13-15, \\
\small Universidad Aut{\' o}noma de Madrid Cantoblanco, Madrid 28049, Spain.\\
\small $^3$ Theoretical Physics Department, Universidad Aut{\' o}noma de Madrid, Madrid 28049, Spain.\\} 

\begin{document} 

\maketitle 

\begin{abstract}
We study the four-gluon scattering amplitude in the high energy limit of QCD written in terms of its conformal expansion. We highlight the need to include both even and odd conformal spin contributions in order to map it to an iterative representation in rapidity and transverse momentum space which we have evaluated numerically. By Fourier expanding in a set of three azimuthal angles, we find a new form for the amplitude in terms of $_4F_3$ hypergeometric functions. An alternative formulation is possible when connecting this Fourier expansion with Bessel kernels studied in analytic number theory. 
\end{abstract}

\section{Introduction}

 There is a renewed interest in studying 
 diffractive hard scattering in hadron-hadron 
 collisions due to the physics program at the LHC~\cite{TOTEM:2021rix,CMS:2017oyi,ATLAS:2014lzu}. When rapidity gaps appear in the final state together with a pair of forward jets well-separated in rapidity~\cite{Mueller:1992pe,Forshaw:2005sx,Chevallier:2009cu,Forshaw:2009fz,Kepka:2010hu,Hentschinski:2014lma,Hentschinski:2014bra,Hentschinski:2014esa,CDF:1997hpy,Motyka:2001zh,Enberg:2001ev} we can attempt to use the BFKL formalism~\cite{Balitsky:1978ic,Kuraev:1977fs,Kuraev:1976ge,Lipatov:1976zz,Fadin:1975cb}  to describe them. There are similar processes where the jets can be replaced by, {\it i.e.}, light vector mesons~\cite{Forshaw:2001pf,Forshaw:1995ax,Enberg:2002zy,Enberg:2003jw,Poludniowski:2003yk,Kotko:2019kma}. This is a rather challenging sector of the strong interaction both from the experimental and theoretical point of view. In this work we address the later, in particular, the structure of the universal quantity present in these types of analysis: the non-forward BFKL gluon Green's function which represents the $t$-channel exchange of a hard Pomeron. In high energy Regge theory this colour singlet exchange  corresponds to the right-most singularity of the $t$-channel partial wave for the elastic gluonic amplitude when evaluated in the complex angular momentum plane. 
 
 The scattering amplitudes relevant to study diffractive  physics simplify in the kinematical region where large rapidity gaps are present, up to the point that they can be resummed to all orders in the QCD coupling accounting, in this way,  for the largest powers in the logarithms with center-of-mass energy dependence. The integral equation generating this resummation, the BFKL equation~\cite{Balitsky:1978ic,Kuraev:1977fs,Kuraev:1976ge,Lipatov:1976zz,Fadin:1975cb}, has a kernel which enjoys two-dimensional conformal invariance in coordinate representation~\cite{Lipatov:1985uk}. This is related to the underlying integrability present in the associated effective field theory~\cite{Lipatov:1993qn,Lipatov:1993yb,Faddeev:1994zg} (see also~\cite{Alfimov:2018cms,Gromov:2015vua,Zhang:2021hra,Hao:2019cfu}). This emerging symmetry is very intriguing from a fundamental point of view and we find it worth investigating from different angles. 
 
In the following we study the BFKL Pomeron singularity, with vacuum quantum numbers,  from two orthogonal standpoints: a numerical Monte Carlo approach and a critical revision of the analytic studies based on the expansion on a conformal basis. The  associated gluon Green's function  carries a representation of the  $SL(2, \mathbb{C})$ group labelled by a real anomalous dimension $\nu$ and an integer conformal spin $n$ via an expansion over eigenfunctions of the M{\" o}bius invariant Hamiltonian. When the Pomeron couples to a certain class of  external states the odd $n$ components do not contribute to the scattering process. These terms are, however, needed to map the Monte Carlo numerical solution for the gluon Green's function and the analytic solutions for non-zero momentum transfer. They might also be present when other external states, in the form of impact factors, will be studied in the future. The relevant step for the  cancellation of a particular conformal spin sector is the integration over some of the three azimuthal angles present in the convolution with those impact factors. It is therefore important to extract this dependence in the Green's function in the most explicit possible form. We also address this point here. It will be shown that this necessarily connects with its $n$ dependence.

Our findings are presented in three sections. In 
the first one we explore the gluon Green's function as the solution to the non-forward BFKL equation in iterative form which we implement numerically using Monte Carlo integration. In the second one we critically review the work of Lipatov~\cite{Lipatov:1985uk} and Navelet-Peschanski~\cite{Navelet:1997xn} for the analytic representation of the previous solution in terms of two-dimensional conformal invariant functions in coordinate space. We offer a solution valid for both even and odd conformal spins which is shown to be equivalent to the one obtained in the previous section by direct numerical integration. We then Fourier expand it to explicitly obtain the intricate dependence on the different azimuthal angles present in the scattering process. A new representation in terms of $ _4F_3$ hypergeometric functions is found. This allows for a simple derivation of the forward limit in the scattering amplitude. We finally sketch an alternative representation, connected to analytic number theory, which brings a novel expansion of the conformal blocks. We conclude with a summary and outlook for future applications of these results. 

\section{Iteration in momentum space}
\label{MCit}

The non-amputated Green's function corresponding to the scattering of four reggeized off-shell gluons follows the Bethe-Salpeter equation~\cite{Lipatov:1985uk}
\begin{eqnarray}
\omega  f_\omega (k_a,k_b,q) &=& \frac{\delta^{(2)} (k_a-k_b)}{k_a^2 (q-k_a)^2}  \nonumber\\
&&\hspace{-2.5cm}
+ \frac{\bar{\alpha}_s}{2 \pi} \int d^2 k
\Bigg\{\bigg[
\frac{ (q-k)^2}{(k-k_a)^2  (q-k_a)^2} +\frac{ k^2}{(k-k_a)^2 k_a^2 } 
- \frac{ q^2}{ k_a^2 (q-k_a)^2}\bigg] 
f_\omega (k,k_b,q)\nonumber\\
&&\hspace{-2.5cm}- \left[\frac{k_a^2}{k^2+(k_a-k)^2}
+\frac{(q-k_a)^2}{(q-k)^2+(k_a-k)^2}\right]  \frac{f_\omega (k_a,k_b,q)}{(k-k_a)^2}\Bigg\}
\label{OriginalEq}
\end{eqnarray}
where the initial condition contains two propagators. To present the first method of solution we shift the integration from $t$-channel to $s$-channel momenta, $k-k_a=l$,
\begin{eqnarray}
\omega  f_\omega (k_a,k_b,q) &=& \frac{\delta^{(2)} (k_a-k_b)}{k_a^2 (q-k_a)^2}  \nonumber\\
&&\hspace{-2.5cm}
+ \frac{\bar{\alpha}_s}{2 \pi} \int d^2 l
\Bigg\{\bigg[\frac{ (q-l-k_a)^2}{l^2  (q-k_a)^2} +\frac{ (l+k_a)^2}{l^2 k_a^2 } 
- \frac{ q^2}{ k_a^2 (q-k_a)^2}\bigg] 
f_\omega (l+k_a,k_b,q)\nonumber\\
&&\hspace{-2.5cm}- \left[\frac{k_a^2}{(l+k_a)^2+l^2}
+\frac{(q-k_a)^2}{(q-k_a-l)^2+l^2}\right]  \frac{f_\omega (k_a,k_b,q)}{l^2}\Bigg\}
\end{eqnarray}
This offers the opportunity to introduce a small cut-off, $\lambda$, not only useful to regularise infrared divergencies but also to motivate the approximation
\begin{eqnarray}
f_\omega (l+k_a,k_b,q) \simeq f_\omega (l+k_a,k_b,q) \theta (l^2-\lambda^2)
+ f_\omega (k_a,k_b,q) \theta (\lambda^2-l^2)
\end{eqnarray}
which allows to operate with a simpler version of the equation in the form
\begin{eqnarray}
f_\omega (k_a,k_b,q) 
&=& \frac{\frac{\delta^{(2)} (k_a-k_b)}{k_a^2 (q-k_a)^2}  
+  \int \frac{ d^2 l}{\pi l^2} \xi (k_a,l,q)
f_\omega (l+k_a,k_b,q) \theta (l^2-\lambda^2)}{\omega   -\omega_\lambda (k_a,q) }
\end{eqnarray}
where
\begin{eqnarray}
\xi (k_a,l,q) &=& \frac{\bar{\alpha}_s}{2} \left(
1 +\frac{ (l+k_a)^2 (k_a-q)^2-q^2 l^2}{ (l+k_a -q)^2k_a^2 }  \right) 
\frac{(l+k_a -q)^2}{(k_a-q)^2}
\end{eqnarray} 

The gluon's Regge trajectory
\begin{eqnarray}
\omega_\lambda (k_a,q) &=& \frac{\bar{\alpha}_s}{2 \pi} \int \frac{d^2 l}{l^2} 
\Bigg\{\bigg[\frac{ (q-l-k_a)^2}{  (q-k_a)^2} +\frac{ (l+k_a)^2}{ k_a^2 } 
- \frac{ q^2 l^2}{ k_a^2 (q-k_a)^2}\bigg] \theta (\lambda^2-l^2)\nonumber\\
&&\hspace{-.5cm}- \frac{k_a^2}{(l+k_a)^2+l^2}
-\frac{(q-k_a)^2}{(q-k_a-l)^2+l^2} \Bigg\}
\end{eqnarray}
can be evaluated noticing that in the forward limit we have
\begin{eqnarray}
\omega_\lambda (k_a,0) = \frac{\bar{\alpha}_s}{\pi} \int \frac{d^2 l}{l^2} 
\Bigg\{\frac{ (l+k_a)^2}{  k_a^2}   \theta (\lambda^2-l^2)- \frac{k_a^2}{(l+k_a)^2+l^2} \Bigg\}
 \simeq - \bar{\alpha}_s \ln{\frac{k_a^2}{\lambda^2}} 
\end{eqnarray}
and, hence,
\begin{eqnarray}
\omega_\lambda (k_a,q) &=& 
\frac{1}{2} \left(\omega_\lambda (k_a,0) +\omega_\lambda (k_a-q,0) \right) -  
\frac{\bar{\alpha}_s}{2 }\frac{ q^2 \lambda^2}{ k_a^2 (q-k_a)^2}
\nonumber\\
&\simeq & 
\frac{1}{2} \left(\omega_\lambda (k_a,0) +\omega_\lambda (k_a-q,0) \right) 
\end{eqnarray}
As long as $\lambda^2 \ll k^2_a ,(q-k_a)^2$ this approximation is valid. 

To write the expression needed for a Monte Carlo evaluation we iterate:
\begin{eqnarray}
f_\omega (k_a,k_b,q) 
&=& \frac{1}{\omega   -\omega_\lambda (k_a,q) }\frac{\delta^{(2)} (k_a-k_b)}{k_a^2 (q-k_a)^2}  \nonumber\\
&&\hspace{-2cm}
+  \int \frac{ d^2 l}{\pi l^2} 
\frac{\theta (l^2-\lambda^2)}{\omega   -\omega_\lambda (k_a,q) }
\frac{\xi (k_a,l,q)}{\omega   -\omega_\lambda (l+k_a,q) }\frac{\delta^{(2)} (l+k_a-k_b)}{(l+k_a)^2 (q-k_a-l)^2} + \dots
\end{eqnarray}
The first term is symmetric under the transformation $\vec{k}_a \leftrightarrow \vec{k}_b$ due to the presence of the delta function. We can integrate the second one,
\begin{eqnarray}
\frac{\theta ((k_a-k_b)^2-\lambda^2)}{\pi (k_a-k_b)^2} 
\frac{1}{(\omega   -\omega_\lambda (k_a,q)) }
\frac{1}{(\omega   -\omega_\lambda (k_b,q) )}\frac{\xi (k_a,k_b-k_a,q)}{k_b^2 (q-k_b)^2}
+ \dots
\end{eqnarray}
and look at the function $\xi$,
\begin{eqnarray}
\frac{\xi (k_a,k_b-k_a,q)}{k_b^2 (q-k_b)^2} = \frac{\bar{\alpha}_s}{2} \left(
\frac{1}{(k_a-q)^2 k_b^2} 
+\frac{ 1}{ (k_b -q)^2k_a^2 } 
-\frac{q^2 (k_b-k_a)^2}{(k_a-q)^2 k_b^2(k_b -q)^2k_a^2} \right)  
\end{eqnarray} 
which also manifests the $\vec{k}_a \leftrightarrow \vec{k}_b$ symmetry. This property holds for any layer in the iteration. We show this in the next term 
\begin{eqnarray}
 \int \frac{ d^2 l_1}{\pi l_1^2}
 \frac{ \xi (k_a,l_1,q)
 \theta (l_1^2-\lambda^2)}{\omega   -\omega_\lambda (k_a,q) }\int \frac{ d^2 l_2}{\pi l_2^2} 
 \frac{ \xi (l_1+k_a,l_2,q)
 \theta (l_2^2-\lambda^2)}{\omega   -\omega_\lambda (l_1+k_a,q) }\nonumber\\
&&\hspace{-7cm} \times
 \frac{1}{\omega   -\omega_\lambda (l_2+l_1+k_a,q) }\frac{\delta^{(2)} (l_2+l_1+k_a-k_b)}{(l_2+l_1+k_a)^2 (q-l_2-l_1-k_a)^2}
\end{eqnarray}
which is more complicated. Making use of the delta functions and the change of variables, $l=l_2-k_b$, 
\begin{eqnarray}
  \int d^2 l \frac{\theta ((l+k_b)^2-\lambda^2) }{\pi (l+k_b)^2} \frac{ \theta ((l +k_a )^2-\lambda^2)}{\pi (l+k_a)^2}
 \frac{ 1 }{\omega   -\omega_\lambda (k_a,q) } \frac{1}{\omega   -\omega_\lambda (k_b,q) }\nonumber\\
&&\hspace{-7cm} \times  
 \frac{ 1}{\omega   -\omega_\lambda (-l,q) }
 \frac{\xi (k_a,-k_a-l,q)\xi (-l,l+k_b,q)}{k_b^2 (q-k_b)^2}
\end{eqnarray}
where
\begin{eqnarray}
 \frac{\xi (k_a,-k_a-l,q)\xi (-l,l+k_b,q)}{k_b^2 (q-k_b)^2} &=&  
 \frac{\bar{\alpha}_s^2}{4} \frac{1}{l^2 (l+q)^2} \nonumber\\
&&\hspace{-6cm} \times \left(\frac{(l+q)^2k_a^2+ l^2 (k_a-q)^2-q^2 (k_a+l)^2}{ k_a^2 (k_a-q)^2}  \right) 
\left(\frac{(k_b -q)^2 l^2+ k_b^2 (l+q)^2-q^2 (l+k_b)^2}{ k_b^2(k_b -q)^2  }  \right) 
\end{eqnarray}
which also respects the symmetry. To work in rapidity space we need
\begin{eqnarray}
f \left({k}_{a}, {k}_{b}, {q}, {Y}\right) &=& \int_{a-i \infty}^{a+i \infty} 
 \frac{d \omega}{2 \pi i}  e^{\omega {Y}} f_{\omega} \left({k}_{a}, {k}_{b}, {q}\right)
 \\
 \int_{a-i \infty}^{a+i \infty} \frac{d \omega}{2 \pi i} e^{\omega Y} \prod_{i=0}^{n} \frac{1}{\omega-\omega_{i}} &=& e^{\omega_{0} Y} \prod_{i=1}^{n} \int_{0}^{y_{i-1}} d y_{i} e^{\omega_{i, i-1} y_{i}}
\end{eqnarray}
We then have, for the first three terms,
\begin{eqnarray}
\int_{a-i \infty}^{a+i \infty} \frac{d \omega}{2 \pi i} e^{\omega Y} f_\omega (k_a,k_b,q) 
&=& e^{\omega_\lambda (k_a,q) Y} 
 \frac{\delta^{(2)} (k_a-k_b)}{k_a^2 (q-k_a)^2}  \nonumber\\
&&\hspace{-5cm}
+ 
 \frac{   \theta ((k_a-k_b)^2-\lambda^2)}{\pi (k_a-k_b)^2}
 \frac{ \xi (k_a,k_b-k_a,q)}{k_b^2 (q-k_b)^2}
\frac{e^{\omega_\lambda (k_a,q) Y}  - e^{\omega_\lambda (k_b,q) Y}}{\omega_\lambda (k_a,q)-\omega_\lambda (k_b,q)}  \nonumber\\
&&\hspace{-5cm}
+ \int d^2 l 
\frac{  \theta ((k_b+l)^2-\lambda^2)}{\pi (k_b+l)^2} 
 \frac{ \theta ((k_a+l)^2-\lambda^2)}{\pi (k_a+l)^2}  
  \frac{\xi (k_a,-k_a-l,q)\xi (-l,l+k_b,q) }{k_b^2 (q-k_b)^2}
\nonumber\\
&&\hspace{-5cm} \times \frac{e^{Y \omega_\lambda (-l,q)} (\omega_\lambda (k_a,q)-
\omega_\lambda (k_b,q))+e^{Y \omega_\lambda (k_a,q)} (\omega_\lambda (k_b,q)-\omega_\lambda (-l,q))+e^{Y
   \omega_\lambda (k_b,q)} (\omega_\lambda (-l,q)-\omega_\lambda (k_a,q))}{(\omega_\lambda (k_a,q)-\omega_\lambda (k_b,q)) (\omega_\lambda (k_a,q)-\omega_\lambda (-l,q))
   (\omega_\lambda (k_b,q)-\omega_\lambda (-l,q))} \nonumber\\
&&\hspace{-5cm}+ \dots
\end{eqnarray}
Finally, the complete iterated representation of the solution reads~\cite{Andersen:2004tt}
\begin{eqnarray}
f \left(k_{a}, k_{b}, q, Y\right) &=&\left(\frac{\lambda^{2}}{k_{a}^{2}} \frac{\lambda^{2}}{\left(k_{a}- q\right)^{2}}\right)^{\frac{\bar{\alpha}_{s}}{2} Y}\Bigg\{ \frac{\delta^{(2)} (k_a-k_b)}{k_a^2 (q-k_a)^2}  +\sum_{n=1}^{\infty} \prod_{i=1}^{n} \nonumber\\
&&\hspace{-3.2cm}  \times \int d^{2} {k}_{i}   \frac{\theta\left({k}_{i}^{2}-\lambda^{2}\right)}{\pi {k}_{i}^{2}}\xi\left({k}_{a}+\sum_{l=1}^{i-1} {k}_{l}, {k}_{i}, {q}\right) 
\int_{0}^{y_{i-1}} \hspace{-.5cm}d y_{i} 
\left(\frac{\left({k}_{a}+\sum_{l=1}^{i-1} {k}_{l}\right)^{2}}{\left({k}_{a}+\sum_{l=1}^{i} {k}_{l}\right)^{2}} \right)^{\frac{\bar{\alpha}_{s}}{2} y_{i}}\nonumber\\
&&\hspace{-3.2cm} \times \left( \frac{\left({k}_{a}+\sum_{l=1}^{i-1} {k}_{l}- {q}\right)^{2}}{\left({k}_{a}+\sum_{l=1}^{i} {k}_{l}- {q}\right)^{2}}\right)^{\frac{\bar{\alpha}_{s}}{2} y_{i}} 
\hspace{-.5cm}
\frac{\delta^{(2)}\left(\sum_{l=1}^{n} {k}_{l}+ {k}_{a}- {k}_{b}\right)}{
\left( {k}_{a}+\sum_{l=1}^{n} {k}_{l}\right)^2 \left({k}_{a}+\sum_{l=1}^{n} {k}_{l}-q\right)^2}\Bigg\}
\label{IterEqn}
\end{eqnarray}
This representation applies at next-to-leading order and in supersymmetric 
field theories~\cite{Andersen:2003an,Andersen:2003wy,Chachamis:2011nz,Chachamis:2012fk,Chachamis:2012qw}, in different $t$-channel color projections~\cite{Andersen:2004uj,Caporale:2013bva}.  It is also important 
for the study of several Reggeon bound states~\cite{Chachamis:2016ejm,Chachamis:2018bys}. In Section~\ref{NumRes} we will show results of the numerical implementation of this equation for the gluon Green's function. The calculation is based on the Monte Carlo analysis of each of the $n$ terms. For a finite value of the coupling and rapidity, numerical convergence is reached after a finite number of iterations. The obtained values will be compared with those extracted from the orthogonal approach based on conformal blocks discussed below.

\section{Conformal representation}

Following the seminal work of Lev Lipatov~\cite{Lipatov:1985uk}, the analytic solution of Eq.~(\ref{OriginalEq}) needs of the eigenfunctions of the $SL(2, \mathbb{C})$  
two-dimensional conformal group in coordinate space (with integer $n$ and real $\nu$)
\begin{eqnarray}
E^{n,\nu} \left(\rho_{1 0}, \rho_{2 0}\right) &=& 
\left(\frac{\rho_{12}}{\rho_{10} \rho_{20}}\right)^{\frac{1-n}{2}+i \nu}
\left(\frac{\rho^*_{12}}{\rho^*_{10} \rho^*_{20}}\right)^{\frac{1+n}{2}+i \nu}
\end{eqnarray}
and eigenvalues
\begin{eqnarray}
\omega (\nu,n)&=& 2 \bar{\alpha}_s \left(\gamma_E - \Re e \,   \psi \left(\frac{|n|+1}{2}+i \nu\right) \right)
\end{eqnarray}
to be written in the (distributional) form
\begin{eqnarray}
\delta^{(2) } (q-q') f_\omega (k,k',q) = \int \prod_{i=1}^2 
\frac{d^2 \rho_i}{(2 \pi)^4}  \prod_{i=1}^2 \frac{d^2 \rho_{i'}}{(2 \pi)^4} 
e^{i k \rho_1 + i (q-k) \rho_2 -i k' \rho_{1'} -i (q'-k') \rho_{2'}}  f_\omega (\rho_1, \rho_2;\rho_{1'}, \rho_{2'})
\end{eqnarray}
where the $n = \pm 1$ cases must be regularized via a principal value prescription. With notation
\begin{eqnarray}
f_\omega (\rho_1, \rho_2;\rho_{1'}, \rho_{2'}) \equiv \sum_{n=-\infty}^\infty \int_{-\infty}^\infty d \nu 
\int d^2 \rho_0 \frac{\left(\nu^2+\frac{n^2}{4}\right) E^{n,\nu} \left(\rho_{1 0}, \rho_{2 0}\right)E^{n,\nu *} \left(\rho_{1' 0}, \rho_{2' 0}\right) }{
(\omega -\omega (\nu,n))
\left(\nu^2+\left(\frac{n+1}{2}\right)^2\right)\left(\nu^2+\left(\frac{n-1}{2}\right)^2\right)} 
\end{eqnarray}

Lipatov~\cite{Lipatov:1985uk} used the simplified mixed representation
\begin{eqnarray}
\frac{1}{(2 \pi)^2} \int d^2 \left(\frac{\rho_1+\rho_2}{2}\right)
e^{-i \vec{q} \, \frac{\vec{\rho}_{11'}+ \vec{\rho}_{22'}}{2}}
f_\omega (\rho_1, \rho_2;\rho_{1'}, \rho_{2'}) &&\nonumber\\
&& \hspace{-6cm} = \frac{|\rho_{12} \rho_{1'2'}|}{16} \sum_{n=-\infty}^\infty 
\int_{-\infty}^\infty d \nu \frac{E_q^{n,\nu} (\rho_{12}) E_q^{n,\nu *} (\rho_{1'2'}) }{ (\omega -\omega (\nu,n))\left(\nu^2+\left(\frac{n+1}{2}\right)^2\right)\left(\nu^2+\left(\frac{n-1}{2}\right)^2\right)}
\end{eqnarray}
where
\begin{eqnarray}
E_q^{n,\nu} (\rho_{12}) &=& 2 
\frac{ \frac{|n|}{2}-i \nu }{\pi 2^{4 i \nu}}
\frac{\Gamma \left( \frac{1+|n|}{2}+i \nu 
\right) \Gamma \left( \frac{|n|}{2}-i \nu \right)}{\Gamma \left( \frac{1+|n|}{2}-i \nu \right)
\Gamma \left( \frac{|n|}{2}+i \nu \right)}
\int \frac{d^2 k}{|\rho_{12}|} e^{i \vec{q} \vec{k}} 
E^{n,\nu} \left(k+\frac{q}{2}, k-\frac{q}{2}\right) 
\end{eqnarray}
Making use of the conformal eigenfunction equations, Navelet and 
Peschanski~\cite{Navelet:1997xn} pointed out the conformal block structure of this expression and proposed its  expansion in terms of products of Bessel functions of the first kind, $J_\mu$ (we will come back to this point in the next section),
\begin{eqnarray}
E_{q}^{n, \nu}(\rho_{12}) &=& \frac{C_{\beta, \tilde{\beta}} J_{\beta}(y) J_{\tilde{\beta}}(y^*)+C_{-\beta,-\tilde{\beta}} J_{-\beta}(y) J_{-\tilde{\beta}}(y^*)}{{q^*}^{\beta}q^{\tilde{\beta}}} 
\end{eqnarray}
where $y= q^* \rho_{12} /4$. They fixed the coefficients in agreement with the small $|q|$ 
boundary condition for $E_{q}^{n, \nu}(\rho)$ calculated by Lipatov~\cite{Lipatov:1985uk}, {\it i.e.}
\begin{eqnarray}
 E_{q}^{n, \nu}(\rho_{12}) &=& {q^*}^{i \nu-\frac{n}{2}} q^{i \nu+\frac{n}{2}} 2^{-6 i \nu} 
 \Gamma\left(\frac{2+|n|}{2}-i \nu\right) 
 \Gamma\left(\frac{2-|n|}{2}-i \nu\right) \nonumber\\  
 &\times& \left[J_{\frac{n}{2}-i \nu}\left(\frac{{q^*} \rho_{12}}{4} \right) J_{-\frac{n}{2}-i \nu}\left(\frac{q {\rho_{12}^*}}{4} \right)-(-1)^{n} J_{-\frac{n}{2}+i \nu}\left(\frac{ {q^*} \rho_{12}}{4}\right) J_{\frac{n}{2}+i \nu}\left(\frac{q {\rho_{12}^*}}{4} \right)\right]  
\end{eqnarray}
The inversion formula 
\begin{eqnarray}
E^{n, \nu}\left(\rho_{10}, \rho_{20}\right) = 
\frac{2^{4 i \nu}}{ \frac{|n|}{2}-i \nu }
\frac{\Gamma \left( \frac{1+|n|}{2}-i \nu \right)
\Gamma \left( \frac{|n|}{2}+i \nu \right)}{\Gamma \left( \frac{1+|n|}{2}+i \nu 
\right) \Gamma \left( \frac{|n|}{2}-i \nu \right)} \frac{\left|\rho_{12}\right|}{8 \pi}
 \int d^{2} \vec{q} \, e^{-i \frac{\vec{q}}{2}\left(\vec{\rho}_{10}+\vec{\rho}_{20}\right)} E_{q}^{n, \nu}\left(\rho_{12}\right)
\end{eqnarray}
is interesting since we can use it to evaluate
\begin{eqnarray}
\int d^2 \vec{\rho}_0 E^{n, \nu}\left(\rho_{10}, \rho_{20}\right) 
E^{n, \nu *}\left(\rho_{1'0}, \rho_{2'0}\right) = 
\frac{\left|\rho_{12}\right| 
\left|\rho_{1'2'}\right|}{16 \left(\nu^2 + \frac{n^2}{4}\right)}
  \int d^{2} \vec{p} \, 
e^{-i \frac{\vec{p}}{2}\left(\vec{\rho}_{11'}+\vec{\rho}_{22'}\right)} 
E_{p}^{n, \nu}\left(\rho_{12}\right)
 E_{p}^{n, \nu *}\left(\rho_{1'2'}\right)
\end{eqnarray}
and, hence,
\begin{eqnarray}
\delta^{(2) } (q-q') f_\omega (k,k',q) &=&   \int 
d^2 \rho_1 d^2 \rho_2  
d^2 \rho_{1'} d^2 \rho_{2'}
e^{i k \rho_1 + i (q - k ) \rho_2 -i k' \rho_{1'} -i (q'-k') \rho_{2'}}  \nonumber\\
&\times& \sum_{n=-\infty}^\infty \int_{-\infty}^\infty
 \frac{ \left|\rho_{12}\right| 
\left|\rho_{1'2'}\right| 
 \int \frac{d^{2} p }{(2 \pi)^8} 
 e^{-i \frac{\vec{p}}{2}\left(\vec{\rho}_{11'}+\vec{\rho}_{22'}\right)} 
E_{p}^{n, \nu}\left(\rho_{12}\right)
 E_{p}^{n, \nu *}\left(\rho_{1'2'}\right)}{16 
 (\omega -\omega (\nu,n))
 \left(\nu^2+\left(\frac{n+1}{2}\right)^2\right)\left(\nu^2+\left(\frac{n-1}{2}\right)^2\right)
}  
\end{eqnarray}

The different set of variables, proposed by Navelet-Peschanski~\cite{Navelet:1997xn} and with a Jacobian $1/2$ in each complex sector,
\begin{eqnarray}
\rho_{1,2} = \frac{1}{2} (b\pm\rho+\sigma ) \, \, \, , \, \, \, 
{\rho}_{1',2'} = \frac{1}{2} (\sigma -b \pm \rho' )
   \end{eqnarray}  
translate this expression into the form
\begin{eqnarray}
 f_\omega (k,k',q) &=&  \frac{1}{2^{10} \pi^4}  
\sum_{n=-\infty}^\infty \int_{-\infty}^\infty  
  d \nu \frac{ \left| \Gamma\left(\frac{2+|n|}{2}+i \nu\right) \Gamma\left(\frac{2-|n|}{2}+ i \nu\right)
\right|^2}{(\omega -\omega (\nu,n))
\left(\nu^2+\left(\frac{n+1}{2}\right)^2\right)\left(\nu^2+\left(\frac{n-1}{2}\right)^2\right)
}        \nonumber\\
&\times& \int d \rho  d \rho^* |\rho|
e^{\frac{i}{2} \rho (k^*- \frac{q^*}{2} )} 
e^{\frac{i}{2} \rho^* (k- \frac{q}{2} )}  \int d \rho'  d {\rho'}^* |\rho'|
 e^{  -\frac{i}{2}  \rho' ( {k'}^*-\frac{{q}^*}{2}  )  } 
e^{  -\frac{i}{2} {\rho'}^* ( k'-\frac{q}{2} ) }   
 \nonumber\\  
 &\times& \left[J_{\frac{n}{2}-i \nu}\left(\frac{{q}^* \rho}{4} \right) J_{-\frac{n}{2}-i \nu}\left(\frac{q {\rho^*}}{4} \right)-(-1)^{n} J_{-\frac{n}{2}+i \nu}\left(\frac{ {q^*} \rho}{4}\right) J_{\frac{n}{2}+i \nu}\left(\frac{q {\rho^*}}{4} \right)\right] \nonumber\\ 
 &\times& \left[J_{\frac{n}{2}+i \nu}\left(\frac{{q} {\rho'}^*}{4} \right) J_{-\frac{n}{2}+i \nu}\left(\frac{q^* {\rho'}}{4} \right)-(-1)^{n} J_{-\frac{n}{2}-i \nu}\left(\frac{ {q} {\rho'}^*}{4}\right) J_{\frac{n}{2}-i \nu}\left(\frac{q^* {\rho'}}{4} \right)\right] 
 \label{FirstExpression}
\end{eqnarray}
This result is different to the one presented by Navelet-Peschanski~\cite{Navelet:1997xn} in 
the odd conformal spins sector. We find that this sector does not cancel and it is needed in order to match the results obtained from the previous section. For the even conformal spin case we find perfect agreement. 

With $\vec{l} = \vec{k}-\frac{\vec{q}}{2}$, $\theta= \theta_\rho 
- \theta_l$, $\theta_l-\theta_q = -\psi$ the integral to be solved now is
\begin{eqnarray}
{\cal I }_{abcd} = \int d \rho  \, d \rho^*
\rho^{\frac{1+\alpha}{2}} {\rho^*}^{\frac{1+\alpha}{2}}
 e^{\frac{i}{2} \rho (k^*- \frac{q^*}{2} )} 
e^{\frac{i}{2} \rho^* (k- \frac{q}{2} )} 
 J_{a \frac{n}{2}+ b i \nu}\left(\frac{{q}^* \rho}{4} \right) 
 J_{c \frac{n}{2} + i d \nu}\left(\frac{q {\rho^*}}{4} \right) &&\nonumber\\
   &&\hspace{-12.5cm}= \frac{2}{i} \int_0^\infty  d |\rho| 
  \int_0^{2 \pi} d \theta  
 e^{i |\rho| |l| \cos{\theta}}   
{|\rho|}^{2+\alpha} 
 J_{a\frac{n}{2}+ b i \nu}\left(\frac{|q| |\rho| e^{i (\theta - \psi)}}{4} \right) 
 J_{c\frac{n}{2}+ d i \nu}\left( \frac{|q| |\rho| e^{-i (\theta - \psi)}}{4}\right) 
\end{eqnarray}
Using
\begin{eqnarray}
J_{\mu}\left( x \right)= \sum_{k=0}^\infty \frac{(-1)^k \left(\frac{x}{2}\right)^{2k +\mu}}{\Gamma(1+k)\Gamma(1+k+\mu)}  \, \, \, \, \, \, , \, \, \, \, \, \, 
\int_{0}^{2 \pi} \frac{d \theta}{2 \pi}  e^{i (x \cos{\left(\theta\right)}- m \theta)} =   
e^{i m \frac{\pi}{2}} J_m (x)
\end{eqnarray}
we obtain
\begin{eqnarray}
{\cal I }_{abcd} &=&   
\sum_{m=0}^\infty \frac{(-1)^m \left(\frac{|q|  e^{-i  \psi}}{8} \right)^{2m +a\frac{n}{2}+ b i \nu}}{\Gamma(1+m)\Gamma(1+m+a\frac{n}{2}+ b i \nu)} 
\sum_{k=0}^\infty \frac{(-1)^k \left( \frac{|q| e^{i  \psi}}{8}\right)^{2k +c\frac{n}{2}+ d i \nu}}{\Gamma(1+k)\Gamma(1+k+c\frac{n}{2}+ d i \nu)} \nonumber\\
&&\hspace{-1cm}\times
\frac{2}{i} \int_0^\infty  d |\rho| \,  {|\rho|}^{2+\alpha+2 (m +k) +(a +c)\frac{n}{2}+ (b + d) i \nu}
  \int_0^{2 \pi} d \theta  
 e^{i |\rho| |l| \cos{\theta}}    e^{i \theta (2(m -k)
 +(a -c)\frac{n}{2}+ (b - d) i \nu)}
\end{eqnarray}

It turns out that $b$ always equals $d$ and this simplifies the calculation. We will need
\begin{eqnarray}
  \int_0^{2 \pi} d \theta  
 e^{i |\rho| |l| \cos{\theta}}    e^{i \theta (2(m -k)
 +(a -c)\frac{n}{2})} = 2 \pi  e^{ i   (k -m+(c -a)\frac{n}{4})\pi} J_{ 2(k -m)
 +(c -a)\frac{n}{2}} (|\rho| |l| )
\end{eqnarray}
in order to obtain
\begin{eqnarray}
{\cal I }_{abcd} &=&   \frac{4 \pi}{i} e^{ i   (c -a)\frac{n}{4}\pi}
\sum_{m=0}^\infty \frac{(-1)^m e^{- i  m \pi}   
\left(\frac{|q|  e^{-i  \psi}}{8} \right)^{2m +a\frac{n}{2}+ b i \nu}}{\Gamma(1+m)\Gamma(1+m+a\frac{n}{2}+ b i \nu)} 
\sum_{k=0}^\infty \frac{(-1)^k e^{ i k \pi}  \left( \frac{|q| e^{i  \psi}}{8}\right)^{2k +c\frac{n}{2}+ d i \nu}}{\Gamma(1+k)\Gamma(1+k+c\frac{n}{2}+ d i \nu)} \nonumber\\
&\times&
 \int_0^\infty  d |\rho| \,  {|\rho|}^{2+\alpha+2 (m +k) +(a +c)\frac{n}{2}+ (b + d) i \nu}  
J_{ 2(k -m)+(c -a)\frac{n}{2}} (|\rho| |l| )
 \end{eqnarray}
 
With $z$ real, $\Re{ (\beta)} < \frac{3}{2}$ and 
$\Re (m + \beta) >0$ we have
\begin{eqnarray}
\int_0^\infty dt \, t^{\beta-1} J_m (z t) &=& \frac{{\rm sgn}(z)^m 
2^{\beta -1}  \Gamma
   \left(\frac{m+\beta }{2}\right) }{
   \left| z\right| ^{\beta } \Gamma
   \left(\frac{m-\beta}{2} +1\right)}\nonumber\\
   &=& \frac{ {\rm sgn}(z)^m}{2 \pi }\left(\frac{2}{\left| z\right|
   }\right)^{\beta } \sin \left(\left(\frac{\beta -m}{2}\right) \pi  \right) 
\Gamma \left(\frac{\beta +m}{2}\right) 
\Gamma \left(\frac{\beta -m}{2}\right)
\end{eqnarray}
In our case  
\begin{eqnarray}
 \int_0^\infty  d t \,  {t}^{2+\alpha+2 (m +k) +(a +c)\frac{n}{2}+ (b + d) i \nu}  
J_{ 2(k -m)+(c -a)\frac{n}{2}} (t |l| )\nonumber\\
   &&\hspace{-8cm} = \frac{1}{2 \pi }\left(\frac{2}{\left| l\right|
   }\right)^{3+\alpha+2 (m +k) +(a +c)\frac{n}{2}+ (b + d) i \nu }  \sin 
   \left(\left(\frac{
   3+\alpha+4 m+an+ (b + d) i \nu }{2}\right) \pi  \right) \nonumber\\
   &&\hspace{-8cm} \times \, 
\Gamma \left(\frac{
   3+\alpha+4 k+c n+ (b + d) i \nu }{2}\right) 
\Gamma \left(\frac{
   3+\alpha+4 m+an+ (b + d) i \nu }{2}\right)
\end{eqnarray}
Therefore,
\begin{eqnarray}
{\cal I }_{abcd} &=&   -\frac{2}{i} e^{ i   (c -a)\frac{n}{4}\pi} 
\left(\frac{2}{\left| l\right|  }\right)^{3+\alpha}
   \left(\frac{q^*}{4 l^*} \right)^{a\frac{n}{2}+ b i \nu}
      \left( \frac{q}{4 l}\right)^{c\frac{n}{2}+ d i \nu}
 \cos  
   \left(\left(\frac{\alpha+an+ (b + d) i \nu }{2}\right) \pi  \right) \nonumber\\
&\times&
\sum_{m=0}^\infty \frac{    
\left(\frac{q^*}{4 l^*} \right)^{2m}\Gamma \left(\frac{
   3+\alpha+4 m+an+ (b + d) i \nu }{2}\right)}{\Gamma(1+m)\Gamma(1+m+a\frac{n}{2}+ b i \nu)}
\sum_{k=0}^\infty \frac{  
\left( \frac{q}{4 l}\right)^{2k } \Gamma \left(\frac{
   3+\alpha+4 k+c n+ (b + d) i \nu }{2}\right)}{\Gamma(1+k)\Gamma(1+k+c\frac{n}{2}+ d i \nu)} 
 \end{eqnarray} 
Setting the regulator to zero, $\alpha = 0$, and performing the two sums we get 
\begin{eqnarray}
{\cal I }_{abcd} &=&   - \frac{2}{i} 
e^{i (c -a)\frac{n}{4} \pi}
\left(\frac{2}{\left| l\right|}\right)^{3  } 
\left(\frac{q^*}{4 l^* } \right)^{a\frac{n}{2}+ b i \nu} 
\left( \frac{q }{4 l }\right)^{c\frac{n}{2}+ d i \nu}
\nonumber\\
&\times&
 \cos  
   \left(\left(\frac{an+ (b + d) i \nu }{2}\right) \pi  \right)
\frac{\Gamma \left(\frac{a n   +3+ i (b+d) \nu}{2} \right) }{\Gamma
   \left(\frac{a n +2}{2} + i b \nu\right)} 
   \frac{\Gamma \left(\frac{c n    +3+ i (b+d) \nu}{2} \right) }{\Gamma \left(\frac{c n
   +2}{2} + i d \nu\right)}\nonumber\\
&\times&
   \, _2F_1\left(\frac{a n+3+i (b+d) \nu}{4},\frac{a
   n +5+i (b+d) \nu}{4};\frac{a n+2}{2}+i b \nu;\frac{{q^*}^2}{4 {l^*}^2}\right)\nonumber\\
&\times&
\, _2F_1\left(\frac{c n+3+
   i (b+d) \nu}{4},\frac{c
   n+5+ i (b+d) \nu}{4};\frac{c n+2}{2}+i d \nu ;\frac{q^2}{4
   l^2}\right)
\end{eqnarray}
Making use of the relation for the hypergeometric function,
\begin{eqnarray}
   \, _2F_1\left(A ,A +\frac{1}{2};\frac{B+1}{2};z^2\right)
&=& (1+z)^{-2 A } \, _2F_1\left(2 A ,\frac{B
   }{2};B ;\frac{2 z}{z+1}\right) 
\end{eqnarray}
and, since $2k=2l+q$ and $b=c=d=-a$:
\begin{eqnarray}
{\cal I }_{a,-a,-a,-a} &=&   - \frac{2^4}{i |k|^3} 
e^{-i a\frac{n}{2} \pi} 
     \left(\frac{k q^*}{k^* q}\right)^{ \frac{a n}{2} }  
        \left(\frac{16|k|^2}{|q|^2}\right)^{ i a \nu }
\nonumber\\
&\times&
 \cos  
   \left(\left(\frac{an-2a  i \nu }{2}\right) \pi  \right)
\frac{\Gamma \left(\frac{a n   +3-2 i a\nu}{2} \right) }{\Gamma
   \left(\frac{a n +2}{2} - i a \nu\right)} 
   \frac{\Gamma \left(\frac{-a n    +3- i 2a \nu}{2} \right) }{\Gamma \left(\frac{-a n
   +2}{2} - i a \nu\right)}\nonumber\\
&\times&
   \, _2F_1\left(\frac{a n+3-i 2a\nu}{2} ,\frac{a n+1-2i a \nu
   }{2};a n+1-2i a \nu ;\frac{q^*}{k^*}\right) \nonumber\\
&\times&
  \, _2F_1\left(\frac{-a n+3-
   i 2a \nu}{2} ,\frac{-a n+1-2 i a \nu
   }{2};-a n+1-2 i a \nu ;\frac{q}{k}\right) 
\end{eqnarray}
where (see Eq.~(\ref{FirstExpression}))
\begin{eqnarray}
{\cal I }_{a,-a,-a,-a} (k,q) = \int d x  \, d y
\, x^{\frac{1}{2}} {y}^{\frac{1}{2}}
 e^{\frac{i}{2} x (k^*- \frac{q^*}{2} )} 
e^{\frac{i}{2} y (k- \frac{q}{2} )} 
 J_{a \frac{n}{2}-a i \nu}\left(x\frac{{q}^* }{4} \right) 
 J_{-a \frac{n}{2} - i a \nu}\left(y \frac{q }{4} \right) 
 \end{eqnarray} 
The terms related to the momentum $k$  in the expression~(\ref{FirstExpression}) read
\begin{eqnarray}
{\cal I }_{+---} (k,q)- (-1)^n {\cal I }_{-+++} (k,q) 
&=&   \frac{2^4 (-i)^{n+1}}{\left| k\right|^3} 
\cos \left(\left(\frac{ n }{2}-    i \nu\right) \pi  \right)
\Bigg\{\left(\frac{k^* q}{k q^*} \right)^{\frac{n}{2}} 
\left( \frac{|q| }{4 |k| }\right)^{ 2 i \nu}
\nonumber\\
&&\hspace{-3cm}\times
\frac{\Gamma \left(\frac{3- n}{2} + i  \nu\right) }{\Gamma
   \left(\frac{2- n}{2} + i  \nu\right)} 
 \, _2F_1\left(\frac{3- n}{2}+i  \nu  ,\frac{1-n}{2}+i   \nu;1-n+i 2  \nu ;\frac{q^*}{k^*}\right) \nonumber\\
&&\hspace{-3cm}
\times    \frac{\Gamma \left(\frac{ 3+n}{2} + i  \nu\right) }{\Gamma \left(\frac{2+ n}{2} + i  \nu\right)} \, _2F_1\left( \frac{3+n}{2}+i  \nu  ,\frac{1+n}{2}+i   \nu;1+n+i 2  \nu ;\frac{q}{k}\right) - {\rm c.c.} \Bigg\} 
\end{eqnarray} 
The two terms related to the momentum $k'$ in~(\ref{FirstExpression}) are the complex conjugated of these:
\begin{eqnarray}
\left({\cal I }_{+---} (k',q)- (-1)^n {\cal I }_{-+++} (k',q)\right)^*
\end{eqnarray} 
Therefore
\begin{eqnarray}
 f (k,k',q,Y) =    \sum_{n=-\infty}^\infty \int_{-\infty}^\infty  
d \nu  \, {\cal G}^{(n,\nu)} (k,q) ({\cal G}^{(n,\nu)} (k',q))^*
e^{\alpha Y \omega (\nu,n)}   
\end{eqnarray} 
with 
\begin{eqnarray}
{\cal G}^{(n,\nu)} (k,q) &=& 
\frac{\cos \left(\left(\frac{ n }{2}+    i \nu\right) \pi  \right) 
\Gamma\left(\frac{2+n}{2}+i \nu\right) 
\Gamma\left(\frac{2-n}{2}+ i \nu\right)
 }{2 \pi^2\left| k\right|^3 \left(\nu +i \frac{(n+1)}{2}\right)
\left(\nu +i \frac{(n-1)}{2}\right)} \Bigg\{\left(\frac{k^* q}{k q^*} \right)^{\frac{n}{2}} 
\left( \frac{|q| }{4 |k| }\right)^{ 2 i \nu}
\nonumber\\
&\times& \frac{\Gamma \left(\frac{3- n}{2} + i  \nu\right) }{\Gamma
   \left(\frac{2- n }{2} + i  \nu\right)} 
 \, _2F_1\left(\frac{3- n}{2}+i  \nu  ,\frac{1-n}{2}+i   \nu;1-n+i 2  \nu ;\frac{q^*}{k^*}\right) \nonumber\\
&\times&    \frac{\Gamma \left(\frac{3+ n}{2} + i  \nu\right) }{\Gamma \left(\frac{2+ n}{2} + i  \nu\right)}\, _2F_1\left( \frac{3+n}{2}+i  \nu  ,\frac{1+n}{2}+i   \nu;1+n+i 2  \nu ;\frac{q}{k}\right) - {\rm c.c.} \Bigg\}
\end{eqnarray} 
In terms of angles:
\begin{eqnarray}
 f (k_a,k_b,q,\theta_a,\theta_b,\theta_q,Y) = 
   \sum_{n=-\infty}^\infty \int_{-\infty}^\infty  
d \nu  \, {\cal G}^{(n,\nu)} (k_a,q,\theta_a,\theta_q) 
({\cal G}^{(n,\nu)} (k_b,q,\theta_b,\theta_q))^*
e^{\alpha Y \omega (\nu,n)}   
\end{eqnarray} 
using 
\begin{eqnarray}
{\cal A}^{(n,\nu)} &=& \frac{\cos \left(\left(\frac{ n }{2}+    i \nu\right) \pi  \right) 
\Gamma\left(\frac{2+n}{2}+i \nu\right) 
\Gamma\left(\frac{2-n}{2}+ i \nu\right)
 }{2 \pi^2 \left(\nu +i \frac{(n+1)}{2}\right)
\left(\nu +i \frac{(n-1)}{2}\right)} 
\end{eqnarray} 
we can extract Fourier components
\begin{eqnarray}
{\cal G}^{(n,\nu)} (k_a,q,\theta_a,\theta_q) &=& \frac{{\cal A}^{(n,\nu)} }{\left| k_a\right|^3} \sum_{k,l=0}^\infty\bigg( 
  e^{-i  (n+l-k)(\theta_a-\theta_q)}
{\cal B}^{(n,\nu)}_{l,k} (|k_a|,|q|) \nonumber\\
&-&   e^{i  (n+l-k)(\theta_a-\theta_q)}
{\cal B}^{(n,-\nu)}_{l,k} (|k_a|,|q|) \bigg)
\end{eqnarray}
where
\begin{eqnarray}
{\cal B}^{(n,\nu)}_{l,k} (|k_a|,|q|) &=& \left( \frac{|q| }{4 |k_a| }\right)^{ 2 i \nu} \frac{|q|^{l+k}}{|k_a|^{l+k}}  \nonumber\\
&\times& \frac{\Gamma (1-n+2 i \nu )}{ \Gamma
   \left(\frac{2-n }{2} + i  \nu\right)\Gamma \left(\frac{1-n}{2}+i \nu \right)  }
\frac{  \Gamma \left(k+\frac{1-n}{2}+i \nu \right) \Gamma
   \left(k+\frac{3-n}{2}+i \nu \right)}{k! \Gamma (k-n+2 i
   \nu +1)} \nonumber\\
&\times& \frac{\Gamma (1+n+2 i \nu )}{  \Gamma \left(\frac{2+n}{2} + i  \nu\right)\Gamma \left(\frac{1+n}{2}+i \nu \right) }
\frac{ \Gamma \left(l+\frac{1+n}{2}+i \nu \right) \Gamma \left(l+\frac{3+n}{2}+i \nu
   \right)}{l!  \Gamma (l+n+2 i \nu +1)}
\end{eqnarray}
 
It is useful to introduce the azimuthal angle  
projection 
\begin{eqnarray}
{\cal D}_m^{(n,\nu)} (|k_a|,|q|) &\equiv& \int_0^{2 \pi} \frac{d \theta_a}{2 \pi} e^{i m (\theta_a-\theta_q)} 
{\cal G}^{(n,\nu)} (k_a,q,\theta_a,\theta_q) 
 \nonumber\\
&=& 
\frac{{\cal A}^{(n,\nu)} }{\left| k_a\right|^3}
   \sum_{k,l=0}^\infty  \bigg(
 \delta_{n+l-k-m,0} {\cal B}^{(n,\nu)}_{l,k} (|k_a|,|q|)
 -  \delta_{n+l-k+m,0} {\cal B}^{(n,-\nu)}_{l,k} (|k_a|,|q|) \bigg)
\end{eqnarray}
which implies the expansion
  \begin{eqnarray}
 f (k_a,k_b,q,\theta_a,\theta_b,\theta_q,Y) &=& 
  \sum_{n=-\infty}^\infty 
 \int_{-\infty}^\infty  d \nu  \, e^{\alpha Y \omega (\nu,n)}  \nonumber\\
&\times& 
 \sum_{M=-\infty}^\infty  {\cal D}_M^{(n,\nu)} (|k_a|,|q|) e^{i M (\theta_q - \theta_a)}
 \sum_{N=-\infty}^\infty  {\cal D}_{-N}^{(n,-\nu)} (|k_b|,|q|) 
  e^{i N (\theta_q - \theta_b)} 
 \end{eqnarray}
 This representation is useful since it makes 
 explicit the azimuthal angle dependence of the Green's function. We can further present this result in an alternative way.  
The eigenvalue of the kernel can be written in the form
\begin{eqnarray}
\omega (\nu,n) = \tilde{\omega} (\nu,n) + (\tilde{\omega} (\nu,n))^* \, \, \, , \, \, \, 
\frac{\tilde{\omega} (\nu,n)}{ \bar{\alpha}_s}  = \psi(1) -\psi \left(\frac{1+|n|}{2}+ i \nu\right) 
\end{eqnarray} 
This allows for
 \begin{eqnarray}
 f (k_a,k_b,q,Y) =  
\sum_{n=-\infty}^\infty \int_{-\infty}^\infty  
d \nu \, {\cal M}^{(n,\nu)}(k_a,q,Y) ({\cal M}^{(n,\nu)}(k_b,q,Y))^*   
\end{eqnarray}  
where
\begin{eqnarray}
{\cal M}^{(n,\nu)}(k_a,q,Y) &=& e^{Y \tilde{\omega} (\nu,n)} 
\frac{\sqrt{{\cal P}^{(n,\nu)}}}{2\pi^2 |k_a|^3}
\bigg(\left(\frac{k_a^* q}{k_a q^*} \right)^{\frac{n}{2}} 
\left( \frac{|q| }{4 |k_a| }\right)^{ 2 i \nu} 
 \nonumber\\
&\times& \frac{\Gamma \left(\frac{3+ n}{2} + i  \nu\right) }{\Gamma \left(\frac{2+ n}{2} + i  \nu\right)}
\, _2F_1\left( \frac{3+n}{2}+i  \nu  ,\frac{1+n}{2}+i   \nu;1+n+i 2  \nu ;\frac{q}{k_a}\right)
\nonumber\\
&\times&   \frac{\Gamma \left(\frac{3- n}{2} + i  \nu\right) }{\Gamma
   \left(\frac{2- n}{2} + i  \nu\right)}
 \, _2F_1\left(\frac{3- n}{2}+i  \nu  ,\frac{1-n}{2}+i   \nu;1-n+i 2  \nu ;\frac{q^*}{k_a^*}\right)  - {\rm c.c.} \bigg)
\end{eqnarray}  
and
\begin{eqnarray}
{\cal P}^{(n,\nu)} &=& \frac{((-1)^{n}+\cosh (2 \pi  \nu )) \left| \Gamma\left(\frac{2+n}{2}+i \nu\right) \Gamma\left(\frac{2-n}{2}+ i \nu\right)
\right|^2}{2 \left(\nu^2+\left(\frac{n+1}{2}\right)^2\right)\left(\nu^2+\left(\frac{n-1}{2}\right)^2\right)} \nonumber\\
&=& \pi ^2 
\frac{ \left| \Gamma\left(\frac{2+n}{2}+i \nu\right) \Gamma\left(\frac{2-n}{2}+ i \nu\right)
\right|^2}{\left|    \Gamma \left(\frac{3+n}{2}+i \nu \right) \Gamma \left(\frac{3-n}{2}+i \nu \right)\right|^2}
\label{simpler}
\end{eqnarray}

The angular dependence can be made more explicit in a similar way to the calculations above, {\it i.e.}
\begin{eqnarray}
{\cal M}^{(n,\nu)}(k_a,q,Y) = e^{Y \tilde{\omega} (\nu,n)} \sqrt{{\cal P}^{(n,\nu)}}
 \sum _{l,k=0}^{\infty } 
\bigg(e^{i (n+l-k) (\theta_q - \theta_a)}  \widetilde{\cal B}_{l,k}^{(n,\nu)}(|k_a|,|q|)&&\nonumber\\
&&\hspace{-8.cm}- e^{-i (n+l-k) (\theta_q - \theta_a)}  \widetilde{\cal B}_{l,k}^{(n,-\nu)}(|k_a|,|q|)\bigg)
\end{eqnarray} 
with $ \widetilde{\cal B}_{l,k}^{(n,\nu)}(|k_a|,|q|) = \frac{{\cal B}_{l,k}^{(n,\nu)}(|k_a|,|q|)}{2\pi^2 |k_a|^3}$. We can again investigate the projection
\begin{eqnarray}
 \widetilde{\cal D}_M^{(n,\nu)}(|k_a|,|q|,Y)
\equiv \int_0^{2 \pi} \frac{d \theta_a}{2 \pi} e^{i M (\theta_a-\theta_q)} 
{\cal M}^{(n,\nu)}(k_a,q,Y)&& \nonumber\\
&&\hspace{-7cm} = e^{Y \tilde{\omega} (\nu,n)} \sqrt{{\cal P}^{(n,\nu)}}   \sum _{l,k=0}^{\infty } \bigg(   \delta_{n+l-M}^k  
\widetilde{\cal B}_{l,k}^{(n,\nu)}(|k_a|,|q|)-  \delta_{n+l+M}^k \widetilde{\cal B}_{l,k}^{(n,-\nu)}(|k_a|,|q|)\bigg)
\end{eqnarray}  
which implies
\begin{eqnarray}
 f (k_a,k_b,q,Y) =  
\sum_{n,M,L=-\infty}^\infty \int_{-\infty}^\infty  
d \nu    \, 
\widetilde{\cal D}_M^{(n,\nu)} (|k_a|,|q|,Y)  e^{-i M (\theta_a-\theta_q)}
  \widetilde{\cal D}_L^{(n,-\nu)} (|k_b|,|q|,Y)  e^{i L (\theta_b-\theta_q)} 
\end{eqnarray}

We can proceed further in the analysis of this expansion:
\begin{eqnarray}
\sum_{M=-\infty}^\infty \widetilde{\cal D}_M^{(n,\nu)} (|k_a|,|q|,Y)  e^{-i M (\theta_a-\theta_q)} = e^{Y \tilde{\omega} (\nu,n)} \sqrt{{\cal P}^{(n,\nu)}} 
 \sum_{S=0}^\infty    \bigg(  e^{-i (S+n) (\theta_a-\theta_q)}
{\cal H}_{S}^{(n,\nu)}(|k_a|,|q|) -   {\rm c.c.}\bigg)
\end{eqnarray}
where
\begin{eqnarray} 
{\cal H}_{S}^{(n,\nu)}(|k_a|,|q|)
&\equiv&  \sum _{T=0}^{\infty } 
\widetilde{\cal B}_{T+S,T}^{(n,\nu)}(|k_a|,|q|) 
~=~  \frac{1}{2\pi^2 |k_a|^3}
\left( \frac{|q| }{4 |k_a| }\right)^{ 2 i \nu} \frac{|q|^{S}}{|k_a|^{S}} \nonumber  \\
 &&\hspace{-2.5cm} \times 
\frac{\Gamma (1+n+2 i \nu)}{\Gamma \left(\frac{2+ n}{2} + i  \nu\right)\Gamma \left(\frac{1+n}{2}+i \nu \right)} 
\frac{\Gamma (1-n+2 i \nu) }{\Gamma \left(\frac{2- n}{2} + i  \nu\right) \Gamma \left(\frac{1-n}{2}+i \nu \right)}\nonumber\\
&&\hspace{-2.5cm} \times \sum _{T=0}^{\infty } \frac{|q|^{2T}}{|k_a|^{2T}} 
 \frac{   \Gamma \left(T+\frac{1-n}{2}+i \nu\right) \Gamma \left(T+\frac{3-n}{2}+i \nu \right)}{
 T!  \, \Gamma (T+1-n+2 i \nu)} \frac{ \Gamma \left(T+S +\frac{1+n}{2}+i \nu\right)
   \Gamma \left(T+S+\frac{3+n}{2}+i \nu \right)}{(T+S)! \,     \Gamma (T+S+1+n+2 i \nu )} \nonumber\\
 &&\hspace{-2.5cm}  =  \frac{1}{2\pi^2 |k_a|^3}
\left( \frac{|q| }{4 |k_a| }\right)^{ 2 i \nu} \frac{|q|^{S}}{|k_a|^{S}}  
\frac{\Gamma (1+n+2 i \nu)  \Gamma \left(\frac{1+n}{2} + S+ i \nu\right) \Gamma \left(\frac{3-n }{2} + i\nu\right)  \Gamma \left(\frac{3+n}{2} + S+ i \nu\right) }{ S!  \, \Gamma (1+n+S+2 i \nu)\Gamma \left(\frac{1+n}{2}+i \nu \right)\Gamma \left(\frac{2+ n}{2} + i  \nu\right)\Gamma \left(\frac{2- n}{2} + i  \nu\right)}  \nonumber\\
&&\hspace{-2.5cm} \times
   \, _4F_3\bigg(\frac{1-n}{2}+i \nu,\frac{3-n}{2}+i \nu,\frac{1+n}{2}+S+i \nu,\frac{3+n}{2}+S+i \nu;\nonumber\\
&&\hspace{4.cm}  S+1,1-n+2 i \nu,1+n+S+2 i \nu;\frac{|q|^2}{|k_a|^2}\bigg) 
\label{4F3Eq}
\end{eqnarray}

Finally, we can offer the representation 
\begin{eqnarray}
f (k_a,k_b,q,Y) &=&  
\sum_{n=-\infty}^\infty \int_{-\infty}^\infty  
d \nu    e^{Y \tilde{\omega} (\nu,n)} \sqrt{{\cal P}^{(n,\nu)}} 
 \sum_{S=0}^\infty    \bigg(  e^{-i (S+n) (\theta_a-\theta_q)}
  {\cal H}_{S}^{(n,\nu)}(|k_a|,|q|) -   {\rm c.c.}\bigg)\nonumber\\
&\times& 
e^{Y \tilde{\omega} (-\nu,n)} \sqrt{{\cal P}^{(n,\nu)}} 
 \sum_{T=0}^\infty    \bigg(  e^{i (T+n) (\theta_b-\theta_q)}
  {\cal H}_{T}^{(n,-\nu)}(|k_b|,|q|) -   {\rm c.c.}\bigg)
  \label{NewConfEq}
\end{eqnarray}
which highlights the role of the conformal spins $n$ in the Fourier expansion in azimuthal angles. 

We can now understand the forward limit in the following way: 
\begin{eqnarray}
f (k_a,k_b,q=0,Y) &=&  \lim_{|q| \to 0}
\sum_{n=-\infty}^\infty  \int_{-\infty}^\infty  d \nu  \,  {\cal M}^{(n,\nu)} (k_a,q,Y)
 ({\cal M}^{(n,\nu)} (k_b,q,Y))^*
\end{eqnarray}
where
\begin{eqnarray}
\lim_{|q| \to 0}{\cal M}^{(n,\nu)}(k_a,q,Y) &=& e^{Y \tilde{\omega} (\nu,n)} 
\frac{\sqrt{{\cal P}^{(n,\nu)}}}{2\pi^2 |k_a|^3} \nonumber\\
&\times& 
\Bigg[\left(\frac{k_a^* q}{k_a q^*} \right)^{\frac{n}{2}} 
\left( \frac{|q| }{4 |k_a| }\right)^{ 2 i \nu} 
 \frac{\Gamma \left(\frac{3+ n}{2} + i  \nu\right) }{\Gamma \left(\frac{2+ n}{2} + i  \nu\right)}
   \frac{\Gamma \left(\frac{3- n}{2} + i  \nu\right) }{\Gamma
   \left(\frac{2- n}{2} + i  \nu\right)}  - {\rm c.c.} \Bigg] \\
 \lim_{|q| \to 0} ({\cal M}^{(n,\nu)}(k_b,q,Y))^* &=& e^{Y \tilde{\omega} (-\nu,n)} 
\frac{\sqrt{{\cal P}^{(n,\nu)}}}{2\pi^2 |k_b|^3} \nonumber\\
&\times& 
\Bigg[\left(\frac{k_b^* q}{k_b q^*} \right)^{-\frac{n}{2}} 
\left( \frac{|q| }{4 |k_b| }\right)^{ -2 i \nu} 
 \frac{\Gamma \left(\frac{3+ n}{2} - i  \nu\right) }{\Gamma \left(\frac{2+ n}{2} - i  \nu\right)}
   \frac{\Gamma \left(\frac{3- n}{2} - i  \nu\right) }{\Gamma
   \left(\frac{2- n}{2} - i  \nu\right)}  - {\rm c.c.} \Bigg]
\end{eqnarray}
Their product carries the prefactor $e^{Y \omega (\nu,n)} \frac{{{\cal P}^{(n,\nu)}}}{4\pi^4 |k_a|^3 |k_b|^3}$ multiplied by the sum of two contributions:
\begin{eqnarray}
- \left(\frac{k_a^* q}{k_a q^*} \frac{k_b^* q}{k_b q^*} \right)^{\frac{n}{2}} 
\left( \frac{|q| }{4 |k_a| } \frac{|q| }{4 |k_b| }\right)^{ 2 i \nu}
\left( \frac{\Gamma \left(\frac{3+ n}{2} + i  \nu\right) }{\Gamma \left(\frac{2+ n}{2} + i  \nu\right)}
   \frac{\Gamma \left(\frac{3- n}{2} + i  \nu\right) }{\Gamma
   \left(\frac{2- n}{2} + i  \nu\right)}   \right)^2  + {\rm c.c.}
\end{eqnarray}
and
\begin{eqnarray}  
 \frac{\Gamma \left(\frac{3+ n}{2} - i  \nu\right) }{\Gamma \left(\frac{2+ n}{2} - i  \nu\right)}
   \frac{\Gamma \left(\frac{3- n}{2} - i  \nu\right) }{\Gamma
   \left(\frac{2- n}{2} - i  \nu\right)} 
 \frac{\Gamma \left(\frac{3+ n}{2} + i  \nu\right) }{\Gamma \left(\frac{2+ n}{2} + i  \nu\right)}
   \frac{\Gamma \left(\frac{3- n}{2} + i  \nu\right) }{\Gamma
   \left(\frac{2- n}{2} + i  \nu\right)} \left(\left(\frac{k_a  k_b^* }{k_a^* k_b } \right)^{\frac{n}{2}} 
      \left(\frac{|k_a|}{|k_b| }\right)^{ 2 i \nu}+{\rm c.c.}\right)
\end{eqnarray}

 The first one generates fast oscillations when $|q| \to 0$ with a zero net value after  integrating over the variable  
$\nu$. The prefactor is symmetric under the change of sign each in $n$ and $\nu$ as it is the prefactor in the second contribution. We can therefore write for $|q| \to 0$,
\begin{eqnarray}
 f (k_a,k_b,q=0,Y) &=&   \sum_{n=-\infty}^\infty  \int_{-\infty}^\infty  
d \nu \,   e^{Y \omega (\nu,n)} \frac{e^{i n (\theta_a-\theta_b)}}{4 \pi^4 |k_a|^3 |k_b|^3}  \left(\frac{|k_a|}{|k_b| }\right)^{ 2 i \nu}\nonumber\\
   &&\hspace{-3.cm} \times   \frac{((-1)^{n}+\cosh (2 \pi  \nu )) 
\Gamma \left(\frac{3+ n}{2} - i  \nu\right)
\Gamma \left(\frac{3- n}{2} - i  \nu\right)
\Gamma \left(\frac{3+ n}{2} + i  \nu\right)
\Gamma \left(\frac{3- n}{2} + i  \nu\right)
}{ \left(\nu^2+\left(\frac{n+1}{2}\right)^2\right)\left(\nu^2+\left(\frac{n-1}{2}\right)^2\right)}\nonumber\\
      &&\hspace{-3cm}=   \sum_{n=-\infty}^\infty  \int_{-\infty}^\infty  
d \nu \,   e^{Y \omega (\nu,n)} \frac{e^{i n (\theta_a-\theta_b)}}{2 \pi^2 |k_a|^3 |k_b|^3} 
      \left(\frac{|k_a|}{|k_b| }\right)^{ 2 i \nu}  
\end{eqnarray}
where we made use of Eq.~(\ref{simpler}). Some of the phenomenology associated to the different conformal spins in the forward limit has been explored in, {\it e.g.},~\cite{SabioVera:2006cza,SabioVera:2007ndx,SabioVera:2007cxj,Angioni:2011wj}.

\section{An alternative representation}

The classical Bessel functions have a long history in the mathematical 
literature.  They play an important role in analytic number theory in the form of 
Bessel kernels. We will use some of the most recent results in this area in the following. 

We start with Eq.~(\ref{FirstExpression}) making use of the relation (\ref{simpler}) and $z= \frac{\rho q^*}{4}, z^*= \frac{\rho^* q}{4}, v= \frac{\rho' q^*}{4}, 
 v^*= \frac{{\rho'}^* q}{4}$ to write
\begin{eqnarray}
 f_\omega (k,k',q) &=&   \frac{2^3}{\pi^2 |q|^6}  
\sum_{n=-\infty}^\infty \int_{-\infty}^\infty  
d \nu \frac{ \left| \Gamma\left(\frac{2+|n|}{2}+i \nu\right) \Gamma\left(\frac{2-|n|}{2}+ i \nu\right)
\right|^2}{(\omega -\omega (\nu,n))((-1)^{n}+\cosh (2 \pi  \nu ))  \left|    \Gamma \left(\frac{3+n}{2}+i \nu \right) \Gamma \left(\frac{3-n}{2}+i \nu \right)\right|^2
}        \nonumber\\
&\times& \int d z   d z^* |z|
e^{i z \left(2\frac{k^*}{q^*}- 1 \right)} e^{i z^*  \left(2\frac{k}{q}- 1 \right)}  
\int  d v   d {v^*}  |v| 
 e^{  - i   v (2 \frac{{k'}^*}{q^*}-1 )  } 
e^{  - i  v^* ( 2 \frac{k'}{q}-1 ) }  
 \nonumber\\  
 &\times&\left[
 {\cal J}_{\frac{i \nu}{2}, -n}(z) -(-1)^{n} 
 {\cal J}_{-\frac{i \nu}{2}, n}(z) \right] \left[
 {\cal J}_{-\frac{i \nu}{2}, n}(v) 
 -(-1)^{n}  {\cal J}_{\frac{i \nu}{2}, -n}(v)  \right] 
\end{eqnarray}  
where
\begin{eqnarray} 
{\cal J}_{\mu, m}(z) &=& J_{-\frac{m}{2}-2 \mu}(z) J_{\frac{m}{2}-2 \mu}(z^*)
\end{eqnarray} 

Now we introduce the following function for even $m$:
\begin{eqnarray}
{\bf J}_{\mu, m}(z) &=&   
\frac{2 \pi^{2}}{\sin (2 \pi \mu)} 
\left({\cal J}_{\mu, m}(4 \pi \sqrt{z})-{\cal J}_{-\mu,-m}(4 \pi \sqrt{z})\right) 
\end{eqnarray}
motivated by the analysis of Bessel kernels developed in~\cite{Qi:2016q} (rank-two case) which implies 
\begin{eqnarray}
 f^{({\rm even~n})}_\omega (k,k',q) &=&   \frac{1}{\pi^6 |q|^6}  
\sum_{{\rm even} \, n} \int_{-\infty}^\infty  
d \nu \frac{\tanh ^2(\pi  \nu )\left| \Gamma\left(\frac{2+|n|}{2}+i \nu\right) \Gamma\left(\frac{2-|n|}{2}+ i \nu\right)
\right|^2}{(\omega -\omega (\nu,n))  \left|    \Gamma \left(\frac{3+n}{2}+i \nu \right) \Gamma \left(\frac{3-n}{2}+i \nu \right)\right|^2
}        \nonumber\\
&\times& \int d z   d z^* |z|
e^{i z \left(2\frac{k^*}{q^*}- 1 \right)} e^{i z^*  \left(2\frac{k}{q}- 1 \right)}  
 {\bf J}_{\frac{i \nu}{2}, -n} \left(\frac{z^2}{16 \pi^2}\right) \nonumber\\
&\times& \int  d v   d {v^*} |v| e^{  - i   v (2 \frac{{k'}^*}{q^*}-1 )  } 
e^{  - i  v^* ( 2 \frac{k'}{q}-1 ) }      
{\bf J}_{\frac{-i \nu}{2}, n} \left(\frac{v^2}{16 \pi^2}\right) 
\end{eqnarray} 
When $m$ is odd the relevant function reads~\cite{Qi:2016q}
\begin{eqnarray}
{\bf J}_{\mu, m}(z) &=&  
 \frac{2 \pi^{2} i}{\cos (2 \pi \mu)}\left({\cal J}_{\mu, m}(4 \pi \sqrt{z})+{\cal J}_{-\mu,-m}(4 \pi \sqrt{z})\right) 
\end{eqnarray}
and, therefore, 
\begin{eqnarray}
 f^{(\rm odd~n)}_\omega (k,k',q) &=&   \frac{-1}{\pi^6 |q|^6}  
\sum_{{\rm odd}\, n} \int_{-\infty}^\infty  
d \nu \frac{\coth ^2(\pi  \nu )\left| \Gamma\left(\frac{2+|n|}{2}+i \nu\right) \Gamma\left(\frac{2-|n|}{2}+ i \nu\right)
\right|^2}{(\omega -\omega (\nu,n))  \left|    \Gamma \left(\frac{3+n}{2}+i \nu \right) \Gamma \left(\frac{3-n}{2}+i \nu \right)\right|^2
}        \nonumber\\
&\times& \int d z   d z^* |z| 
e^{i z \left(2\frac{k^*}{q^*}- 1 \right)} e^{i z^*  \left(2\frac{k}{q}- 1 \right)}  
{\bf J}_{\frac{i \nu}{2}, -n} \left(\frac{z^2}{16 \pi^2}\right)  \nonumber\\
&\times&
\int  d v   d {v^*} |v| e^{  - i   v (2 \frac{{k'}^*}{q^*}-1 )  } 
e^{  - i  v^* ( 2 \frac{k'}{q}-1 ) }  
 {\bf J}_{-\frac{i \nu}{2}, n} \left(\frac{v^2}{16 \pi^2}\right)
\end{eqnarray} 
The function ${\bf J}_{\mu, m}(z)$ fulfils the equations
\begin{eqnarray}
\left(z^{2} \frac{\partial^{2}}{\partial z^{2}} + z \frac{\partial}{\partial z}+z^{2}
- \left(\frac{n}{2}-i \nu \right)^{2}
\right) {\bf J}_{\frac{i \nu}{2}, -n} \left(\frac{z^{2}}{16 \pi^{2}}\right) &=&0 \\
\left({z^*}^{2} \frac{\partial^{2}}{\partial {z^*}^{2}} + z^* \frac{\partial}{\partial z^*}+ {z^*}^{2}
- \left(\frac{n}{2}+i \nu \right)^{2}
\right) {\bf J}_{\frac{i \nu}{2}, -n} \left(\frac{z^{2}}{16 \pi^{2}}\right)&=&0  
\end{eqnarray}
Both results can be combined in the form
\begin{eqnarray}
 f_\omega (k,k',q) &=&   \frac{-1}{2\pi^5 |q|^6}  
\sum_{n=-\infty}^\infty \int_{-\infty}^\infty  
d \nu \frac{ \frac{4^{2 i \nu } \Gamma \left(\frac{(-1)^n+1}{4} +i \nu\right)^2}{\Gamma (2 i \nu )^2 \Gamma
   \left(\frac{(-1)^n+3}{4}- i \nu \right)^2}
\frac{\left| \Gamma\left(\frac{2+|n|}{2}+i \nu\right) \Gamma\left(\frac{2-|n|}{2}+ i \nu\right)
\right|^2}{\left|    \Gamma \left(\frac{3+n}{2}+i \nu \right) \Gamma \left(\frac{3-n}{2}+i \nu \right)\right|^2}}{(\omega -\omega (\nu,n))((-1)^{n}+\cosh (2 \pi  \nu ))  
}        \nonumber\\
&\times& \int d z   d z^* |z|
e^{i z \left(2\frac{k^*}{q^*}- 1 \right)} e^{i z^*  \left(2\frac{k}{q}- 1 \right)}  
{\bf J}_{\frac{i \nu}{2}, -n} \left(\frac{z^2}{16 \pi^2}\right)
 \nonumber\\  
 &\times& \int  d v   d {v^*}  |v| 
 e^{  - i   v (2 \frac{{k'}^*}{q^*}-1 )  } 
e^{  - i  v^* ( 2 \frac{k'}{q}-1 ) }  
{\bf J}_{-\frac{i \nu}{2}, n} \left(\frac{v^2}{16 \pi^2}\right)
\end{eqnarray} 

In~\cite{BruggemanMotohashi:2003bm}  an interesting integral representation was found by Bruggeman and 
Motohashi that, for 
$z= x e^{i \phi}$ with real $x$ and $\phi$, we can write in the form
\begin{eqnarray}
{\bf J}_{\frac{i \nu}{2}, -n} \left(\frac{z^2}{16 \pi^2}\right)   &=&
\frac{2 \pi}{ i^{n}} \Bigg(\int_{0}^{1} {y}^{i  \nu-1}  
\left(\frac{y e^{ i \phi}
+e^{- i \phi}}{y e^{ -i \phi}
+e^{ i \phi}} \right)^{\frac{n}{2}}
 J_{-n}\left( \frac{x}{\sqrt{y}} \left|y e^{ i \phi}+e^{- i \phi}\right|\right) dy  + {\rm c.c.}\Bigg)
 \nonumber\\ 
 &&\hspace{-2cm}=~ 2 \pi \left(\frac{2}{x i}\right)^{n}  
 \int_{0}^{1}  dy \sum_{k=0}^\infty \sum_{s=0}^k  \sum_{t=0}^{k-n}  
\frac{\left(-\frac{x^2}{4}\right)^{k } \left(e^{ i (2(s-t)-n) \phi}  {y}^{s+t-k-1+\frac{n}{2}+i  \nu}     + {\rm c.c.}\right)}{s! t! \Gamma(1+k-s) 
\Gamma(1+k-n-t)}
\end{eqnarray}
The relevant integral can then be expressed as a 
Fourier expansion in $\theta_k - \theta_q$, {\it i.e.}
\begin{eqnarray}
\int d z   d z^*
e^{i z \left(2\frac{k^*}{q^*}- 1 \right)} e^{i z^*  \left(2\frac{k}{q}- 1 \right)}  
z^{\frac{1}{2}} {z^*}^{\frac{1}{2}} {\bf J}_{\frac{i \nu}{2}, -n} \left(\frac{z^2}{16 \pi^2}\right)  &=& \sum_{m=-\infty}^\infty e^{i m (\theta_k - \theta_q)}  
{\cal C}_{n,\nu}^{(m)} \left(\frac{|k|}{|q|}\right)
\label{NewRep}
  \end{eqnarray}   
where
\begin{eqnarray}
{\cal C}_{n,\nu}^{(m)} \left(\frac{|k|}{|q|}\right) &=& \frac{2^{n+3}  \pi^2}{i}    \int_0^\infty  d x    \int_{0}^{1}  dy  \sum_{k=0}^\infty \sum_{s=0}^k  \sum_{t=0}^{k-n} 
\frac{ (- 1)^{k+m+s+t } \, {y}^{s+t-k-1}  \, 
x^{2+2k-n} J_m \left(4 x \frac{|k|}{|q|}\right) }{4^k s! t! \Gamma(1+k-s) 
\Gamma(1+k-n-t)}\nonumber\\ 
&\times&  
 \bigg( 	{y}^{\frac{n}{2}+i  \nu}    J_{m+n-2(s-t)} (-2 x)  
	+ (-1)^{n} {y}^{\frac{n}{2}-i  \nu}  J_{m-n+2(s-t)} (-2 x) \bigg)  
\end{eqnarray}
This expression belongs to the class of so-called ${\bf K}$-transforms for continuous and compactly supported functions on the complex plane used in 
Lemma 2.1 of~\cite{Watt:2013w}. 

\section{Numerical results and comparison}
\label{NumRes}

In the following brief section we numerically evaluate the gluon Green's function both using the Monte Carlo iteration explained in Section~\ref{MCit} and the expressions stemming from the conformal block techniques obtained in the previous two sections. 

We will make the comparison for a finite set of plots which will highlight different dependences on the transverse momenta, azimuthal angles and rapidity present in the four-point non-forward scattering amplitude. The relevant variables are $|\vec{k}_a|, |\vec{k}_b|, |\vec{q}|, \theta_a, \theta_b, \theta_q$ and $Y$. To present our numerical results we have chosen, without loss of generality, 
$|\vec{k}_a| = 17$ GeV, $|\vec{k}_b| = 31$ GeV, 
$\theta_a = \frac{\pi}{3}$, $\theta_b = \frac{9 \pi}{14}$, 
$Y=4$ and $\bar{\alpha}_s=0.2$.

\begin{figure}
\begin{center}
\vspace{-5cm}
\includegraphics[width=13cm,angle=0]{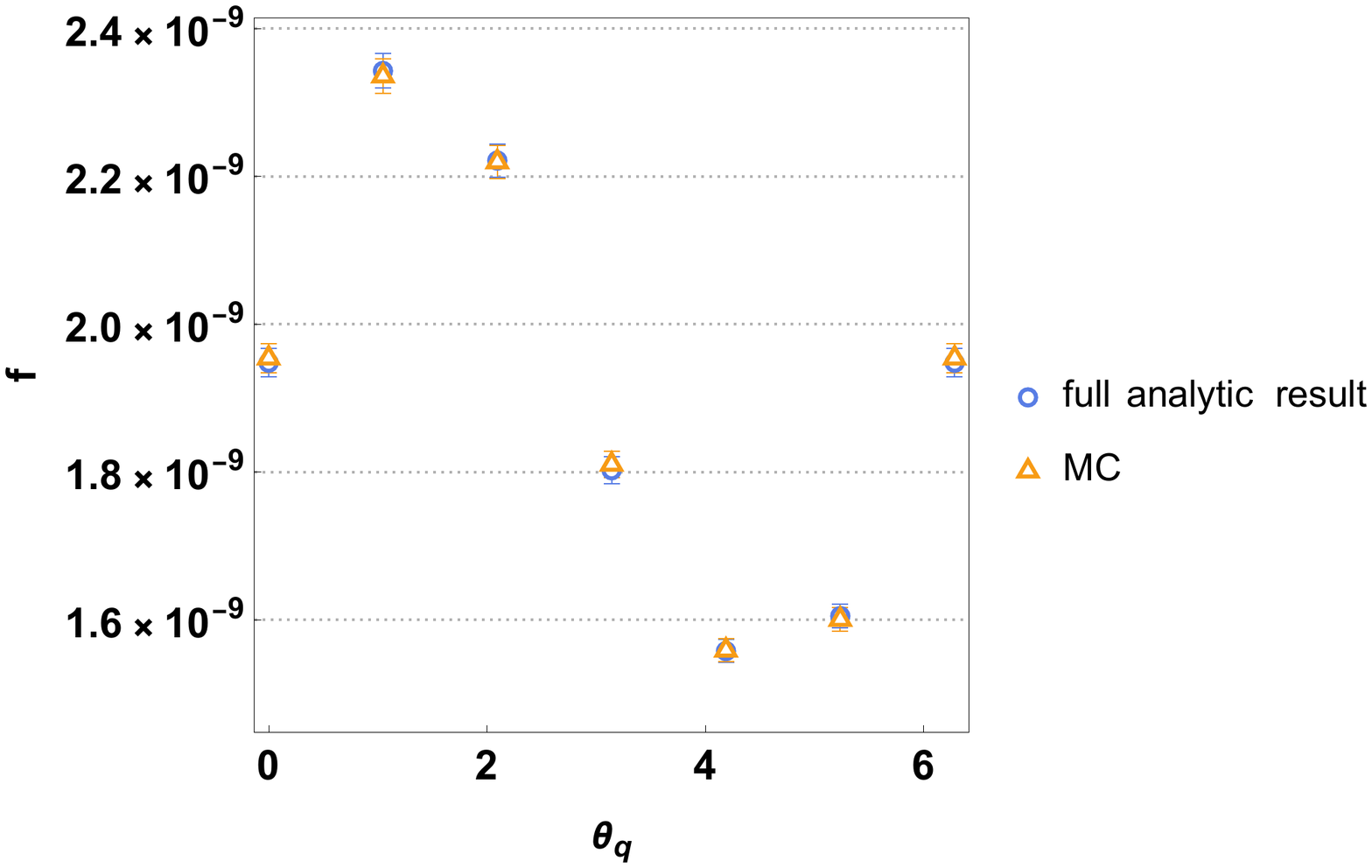}
\vspace{-5cm}
\caption{Gluon Green's function dependence on $\theta_q$ for fixed 
$|\vec{k}_a|, |\vec{k}_b|, |\vec{q}|, \theta_a, \theta_b$ and $Y$.}
\label{Thetaq1A}
\end{center}
\end{figure}

\begin{figure}
\begin{center}
\vspace{-5cm}
\includegraphics[width=13cm,angle=0]{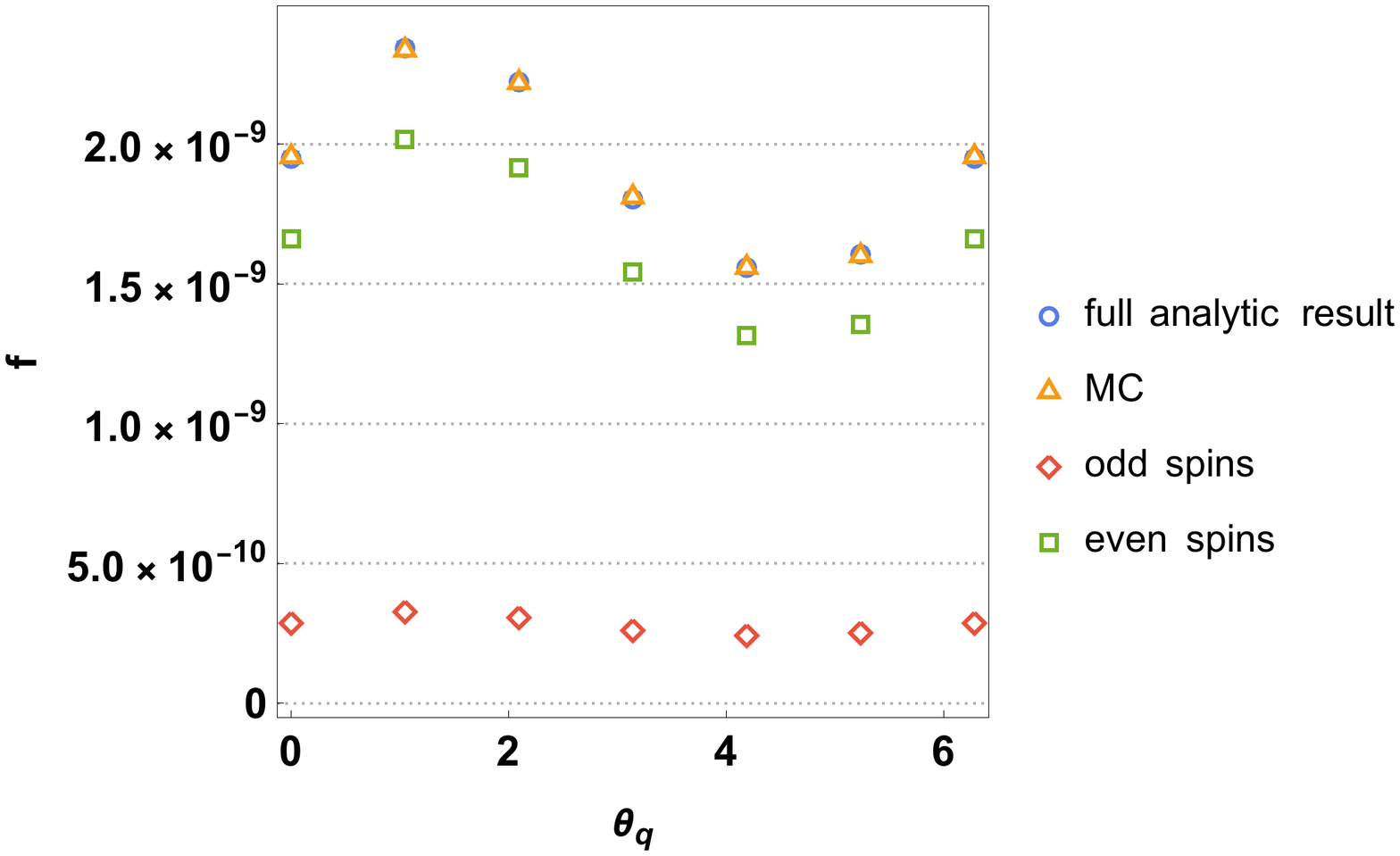}
\vspace{-5cm}
\caption{Gluon Green's function dependence on $\theta_q$ for fixed 
$|\vec{k}_a|, |\vec{k}_b|, |\vec{q}|, \theta_a, \theta_b$ and $Y$.}
\label{Thetaq1B}
\end{center}
\end{figure}

\begin{figure}
\begin{center}
\vspace{-5cm}
\includegraphics[width=12cm,angle=0]{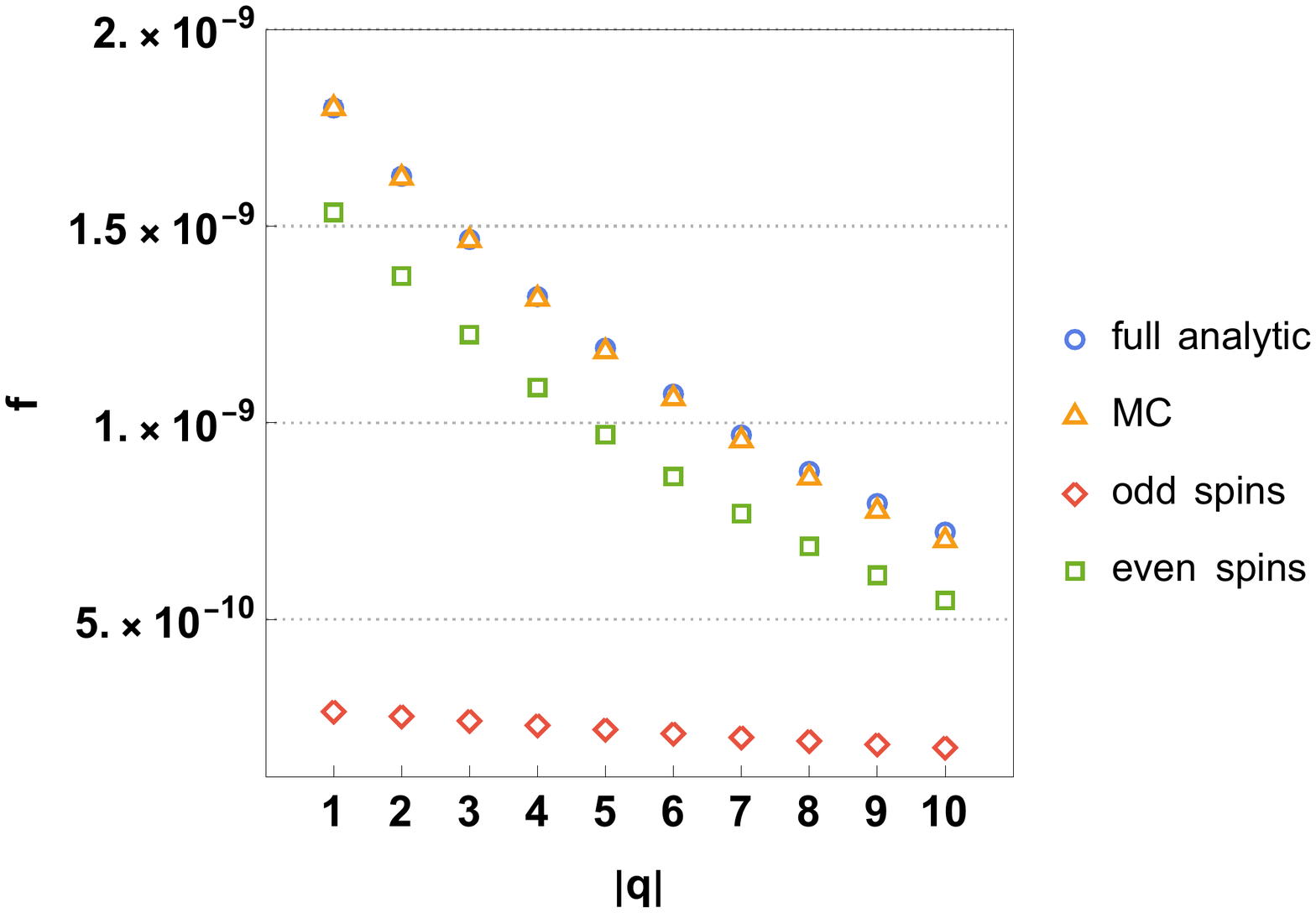}
\vspace{-4.cm}
\caption{Gluon Green's function dependence on $|q|$ for fixed 
$|\vec{k}_a|, |\vec{k}_b|, \theta_a, \theta_b, \theta_q$ and $Y$. In order to match the MC results both odd and even conformal spins must be considered.}
\label{Q1}
\end{center}
\end{figure}

We start by fixing the values of all the variables as above together with $|q|=2$ GeV and studying the effect of the variation on $\theta_q$  from 0 to $2 \pi$. This is shown in Fig.~\ref{Thetaq1A} where we find, within the very small numerical uncertainties, perfect agreement between the analytic and Monte Carlo expressions in the full range of the azimuthal angle present in the two-dimensional momentum transfer vector $\vec{q}$. 
It is now instructive to split the analytic result in two parts: the contributions from all even and all odd conformal spins. As explained in detail in the previous sections, it is found that, contrary to the results of~\cite{Navelet:1997xn}, the 
latter do not cancel and are numerically important to generate the correct gluon Green's function. We highlight this point in 
Fig.~\ref{Thetaq1B} where it is clear that the odd spins dependence is mandatory to reproduce the Monte Carlo iteration results. Note that the dominant conformal spin is always $n=0$ in any case. 
We obtain the same agreement when studying the $|q|$ dependence of the Green's function in Fig.~\ref{Q1} where 
we have fixed $\theta_q= \frac{17 \pi}{10}$. The need of including the odd $n$ sector is manifest in this case as well. 

From our numerical studies we have explicitly shown the validity of the analytic  expressions presented in this work. In particular, we confirm the need of including all (odd and even) conformal spins in the conformal block expressions in order to completely describe the scattering amplitude. We will study the interplay of the Fourier expansions on different azimuthal angles here discussed with the possible types of impact factors in future works. 

\section{Conclusions and outlook}

We have studied  the BFKL Pomeron singularity (the QCD vacuum singularity of the $t$-channel partial wave)  in terms of the non-forward four-reggeized gluon scattering amplitude both with a Monte Carlo approach (Eq.~(\ref{IterEqn})) and  analytically by means of an expansion on a two-dimensional conformal basis. It has been shown that the contributions from all conformal spins, even and odd, are  needed in the latter in order to match the results in the former. We believe this will be relevant for upcoming applications of the BFKL formalism to the future LHC physics program in hard diffraction since one might foresee new types of couplings of external states to the universal BFKL Green's function which might not suppress certain $SL(2, \mathbb{C})$ spin sectors as it happens in, {\it e.g.} the production of a pair of well-separated in rapidity jets with a large rapidity gap in between.

In order to pinpoint this fact in the analytic expressions, we have revisited the work of Lipatov~\cite{Lipatov:1985uk} and Navelet-Peschanski~\cite{Navelet:1997xn} and introduce a Fourier expansion over 
the set of three azimuthal angles present in the amplitude (Eq.~(\ref{NewConfEq})). A novel representation based on $_4F_3$ hypergeometric functions hence arises (Eq.~(\ref{4F3Eq})). The forward limit can then be understood  in simple terms. 

In recent years, mathematicians have developed new tools to investigate different 
representations of the two-dimensional conformal group which are related to 
some of the results here discussed. In particular, in the last section, we have briefly presented an interesting alternative form (Eq.~(\ref{NewRep})) for the conformal blocks present in the BFKL framework inspired by recent results from the mathematical literature devoted to the study of analytic number theory.  

The results here presented have been crossed checked numerically both with a Monte Carlo integration evaluation directly in transverse momentum space and the implementation of the analytic formulae in the form of sums over conformal spins and integration over the anomalous dimension characteristic 
of the $SL(2, \mathbb{C})$ group. The interplay of our findings with different classes of possible impact factors is left for further analysis.

\section*{Acknowledgements}

The work of ASV  is partially supported by the Spanish Research Agency (Agencia Estatal de Investigaci{\'o}n) through the Grant IFT Centro de Excelencia Severo Ochoa No CEX2020-001007-S, funded by MCIN/AEI/10.13039/501100011033 and the Spanish Ministry of Science and Innovation grant PID2019-110058GB-C21/ C22. It has also received funding from the European Union’s Horizon 2020 research
and innovation programme under grant agreement No. 824093. The work of GC was supported by the Funda\c{c}{\~ a}o para a Ci{\^ e}ncia e a Tecnologia (Portugal) under project CERN/FIS-PAR/0024/2019 and contract ‘Investigador FCT - Individual Call/03216/2017’.

\end{document}